\documentclass[twocolumn]{aastex63}

\received{September 26, 2019}
\accepted{by ApJ Letters on November 11, 2019}

\begin{document}

\title{J0453+1559: a neutron star--white dwarf binary from a thermonuclear electron-capture supernova?}
\shorttitle{Is J0453+1559 a neutron star--white dwarf binary?}

\email{tauris@phys.au.dk}
\email{thj@mpa-garching.mpg.de}

\author[0000-0002-3865-7265]{Thomas~M.~Tauris}
\affiliation{Aarhus Institute of Advanced Studies (AIAS), Aarhus University, H{\o}egh-Guldbergs~Gade~6B, 8000~Aarhus~C, Denmark}
\affiliation{Department of Physics and Astronomy, Aarhus University, Ny Munkegade 120, 8000~Aarhus~C, Denmark}

\author[0000-0002-0831-3330]{Hans-Thomas~Janka}
\affiliation{Max Planck Institute for Astrophysics, Karl-Schwarzschild-Str.~1, 85748 Garching, Germany}

\shortauthors{T.~M.~Tauris and H.-Th.~Janka}

\begin{abstract}
The compact binary radio pulsar system J0453+1559 \citep{msf+15}
consists of a recycled pulsar as primary component of $1.559(5)\,M_\odot$
and an unseen companion star of 1.174(4)\,$M_\odot$. 
Because of the relatively large orbital eccentricity of $e = 0.1125$, 
it was argued that the companion is a neutron star, making it 
the neutron star with the lowest accurately determined mass to date. 
However, a direct observational determination of the nature 
of the companion is currently not feasible. Moreover, state-of-the-art stellar evolution and 
supernova modeling are contradictive concerning the possibility to produce such a low-mass neutron star remnant.
Here we challenge the neutron star interpretation by reasoning that the lower-mass component 
could instead be a white dwarf born in a thermonuclear electron-capture supernova
(tECSN) event, in which oxygen-neon deflagration in the degenerate stellar
core of an ultra-stripped progenitor ejects several 0.1$\,M_\odot$ of matter
and leaves a bound ONeFe white dwarf as the second-formed compact remnant. 
We determine the ejecta mass and remnant kick needed in this scenario to
explain the properties of PSR~J0453+1559 by a neutron star--white dwarf system.
More work on tECSNe is needed to assess the viability of this scenario.
\end{abstract}

\keywords{stars: evolution --- stars: neutron --- binaries: general --- 
white dwarfs --- supernovae: general --- pulsars: individual: J0453+1559}

\section{Introduction}\label{sec:intro}
The precise measurements of neutron star (NS) masses play a crucial role in modern astrophysics for many reasons. Their upper limit, currently about $2.0\,M_\odot$ \citep{afw+13,cfr+19}, can be used to constrain the equation-of-state (EoS) of high-density nuclear matter \citep{lp16,of16}, as well as for improving our understanding of the final stages of massive star evolution \citep{lan12} and the subsequent supernova (SN) explosion \citep{jan12,suk+16}. The lower limit of NS masses, however, has received much less attention in the literature until recently, although it has similarly important consequences for improving our knowledge, in particular on SNe and stellar evolution near the lower-mass end of exploding stars \citep[e.g.][]{mth+19} and on the nuclear EoS of NSs, if the baryonic core mass of its progenitor were known \citep[e.g.,][]{klaehn+06}.

The binary radio pulsar J0453+1559 was discovered by \citet{msf+15}. It is a mildly recycled 45~ms pulsar in a 4.07~days orbit with an unseen companion star. The mass of the $1.559(5)\,M_\odot$ pulsar and its $1.174(4)\,M_\odot$ companion star were measured to high accuracy from detections of post-Keplerian parameters: the rate of advance of periastron and the Shapiro delay.
The measured spin period derivative of $\dot{P} = 1.86\times 10^{-19}$ and the estimated surface magnetic field of about $3\times 10^9\,{\rm G}$, strongly suggest that this pulsar was mildly recycled by the accretion of matter from the progenitor of the companion star \citep[e.g.][]{tv06}. 

\begin{figure}[]
\begin{center}
\mbox{\includegraphics[width=0.7\columnwidth, angle=-90]{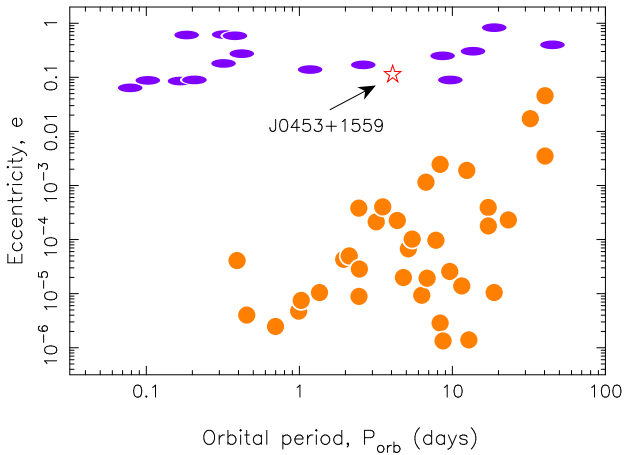}}
  \caption{
    Eccentricity vs.\ orbital period for Galactic disk pulsars with NS companions (purple ellipses) and for recycled pulsars with massive (CO~or~ONeMg)~WD companions (orange circles). The position of PSR~J0453+1559 is marked with a red star. Data from ATNF Pulsar Catalogue in Sept.~2019 \citep{mhth05}.
    }
\label{fig:ecc}
\end{center}
\end{figure}

The double NS nature of this binary was concluded from the measured orbital eccentricity of the system ($e=0.1125$), which is an expected relic of a second SN in the system. As seen in Figure~\ref{fig:ecc}, the eccentricity of PSR~J0453+1559 is indeed typical
among the known population of double NS systems in the Galactic disk. On the contrary, the eccentricities of recycled pulsars with massive white dwarf (CO or ONeMg~WD) companions are much smaller, at least by a factor of 100. The reason for this is that the circularization process of binaries, arising from the strong tidal torques during mass transfer, will leave behind almost perfectly circular systems with very small eccentricities. In systems where this mass-transfer epoch is followed by a SN explosion, the eccentricity instantaneously increases to values typically between $0.1-1$, due to the sudden mass loss and a potential kick added to the newborn NS \citep{hil83}. It is therefore clear that {\em if} PSR~J0453+1559 is not a double NS system, then special circumstances are needed to produce the observed eccentricity.
We notice that there are several known WDs in binaries which have masses $>1.1\,M_\odot$ \citep[e.g.][]{mlt+11,btb+15}. No optical companion has been found at the position of PSR~J0453+1559 \citep{msf+15}. However, a detection is not expected even if the system hosts a massive WD, given the distance (1.1~kpc) and rapid cooling of such a massive WD.

Explaining the origin of a $1.17\,M_\odot$ NS from stellar core-collapse SNe is not
straightforward, and it is evident that stellar evolution modeling for $M_\mathrm{ZAMS} \le 11\,M_\odot$
remains a challenging task \citep{woo15}.
Super-AGB stars are the lowest-mass progenitors whose degenerate cores can become unstable to gravitational collapse. Their ONeMg cores, however, 
have a mass of $\sim\!1.35-1.36\,M_\odot$ \citep{zha+19} and collapse to NSs of
similar baryonic mass \citep{kit+06,jan+08,hue+10,fis+10}, 
corresponding to gravitational masses between $\sim$1.22\,$M_\odot$ and
$\sim\!1.24\,M_\odot$ \citep[via equation~36 of][for NS radii of 11--12\,km]{lat01}.
Although some investigations of stars with zero-age main-sequence (ZAMS) masses of 
$9.35\,M_\odot \le M_\mathrm{ZAMS} \le 9.75\,M_\odot$, which possess 
low-mass CO-cores and Fe-cores of less than 1.3\,$M_\odot$ \citep{suwa+18}, 
as well as studies of stars in the mass range 
$M_\mathrm{ZAMS} < 12\,M_\odot$ \citep{mueller+16}, have determined possible progenitors of $M_\mathrm{NS} < 1.2\,M_\odot$ NSs,
the large population of such low-mass NSs predicted by the 
parametrized explosion models of \citet{mueller+16} seems incompatible with observations
\citep{ant+16} and may point to a problem. Furthermore, SN explosion simulations based on 
other state-of-the-art sets of progenitors do not support the possibility of NS formation
with masses below $\sim\!1.20\,M_\odot$, see \citet{suk+16} and \citet{burrows+19} for single stars and \citet{ewsj19} and \citet{mth+19} for progenitors in binaries.

Therefore, it is a viable question whether the pulsar companion of J0453+1559 could possibly 
be a WD instead of a NS. Here in this {\em Letter}, we propose that the recently
advocated existence of so-called thermonuclear electron-capture SNe (tECSNe), i.e.\
incomplete explosions of degenerate ONeMg cores by oxygen deflagration leaving 
behind bound ONeFe~WD remnants \citep{jrp+16,jon+19,kir+19}, might offer a formation
scenario of J0453+1559 as a NS--WD binary. This model has the advantage that it does 
not invoke the difficult production of a NS of only 1.17\,$M_\odot$. At the same 
time, it can explain the observed eccentricity, arising from explosive mass loss 
of a few $0.1\,M_\odot$.

\section{Results}\label{sec:results}
For a symmetric SN, the relation between the post-SN orbital period ($P_{\rm orb}$) and the pre-SN orbital period ($P_{\rm orb,0}$) is simply given by \citep{bv91}: $P_{\rm orb}=P_{\rm orb,0}\,\mu/(2\mu-1)^{3/2}$, where $\mu =(M_{\rm rem}+M_{\rm comp})/(M_{\rm He}+M_{\rm comp})$ is the ratio between the total system mass after and before the SN. The post-SN eccentricity is given by $e=(1-\mu)/\mu$. Here the pre-SN mass of the exploding star is denoted by $M_{\rm He}$; its (gravitational) remnant mass is denoted by $M_{\rm rem}$, and $M_{\rm comp}$ is the mass of the companion star.
In the case of PSR~J0453+1559, we have $M_{\rm comp}=1.559\,M_\odot$ and $M_{\rm rem}=1.174\,M_\odot$ \citep{msf+15}. Thus for a symmetric SN (i.e.\ without any kick, $w=0$), we find $M_{\rm He}=1.481\,M_\odot$ in order to achieve $e=0.1125$.

\begin{figure}[]
\begin{center}
\mbox{\includegraphics[width=0.7\columnwidth, angle=-90]{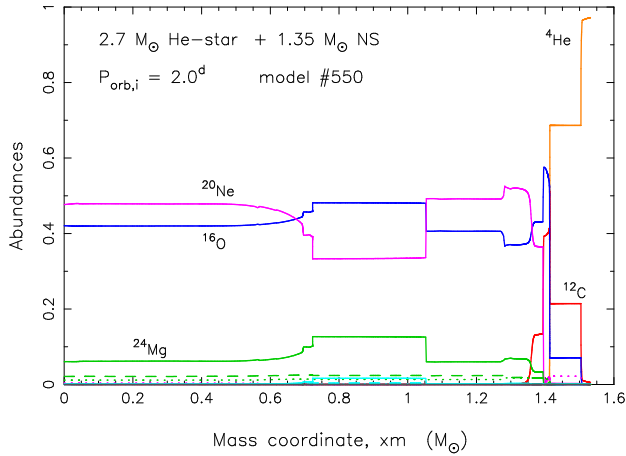}}
  \caption{
     Final chemical abundance structure of an evolved $2.7\,M_\odot$ helium star prior to a potential tECSN event. Here a calculation is shown for a case with an accreting NS of $1.35\,M_\odot$ in an initial orbit of 2.0~days \citep{tlp15}. After mass transfer, the final mass of the exploding star is $1.53\,M_\odot$, with a metal core of $1.42\,M_\odot$. The total amount of Mg is $0.159\,M_\odot$, of which $0.109\,M_\odot$ is $^{24}{\rm Mg}$.
    }
\label{fig:profile}
\end{center}
\end{figure}

Assuming that the threshold mass for undergoing a tECSN is $\sim 1.39\,M_\odot$ \citep{jrp+16}, we would need a progenitor star with a $\sim 1.39\,M_\odot$ metal core and an envelope of $\sim 0.09\,M_\odot$ for this scenario to work for a symmetric SN. Indeed, such (ultra-stripped) SN progenitor models in close binaries where the first-formed compact object is a NS were studied in detail by \citet{tlp15}. In their Table~1, we find such ultra-stripped progenitors for binary models calculated from initial helium star masses of 2.6 and $2.7\,M_\odot$, and an initial orbital period of 2.0~days. Furthermore, these two models have final core masses of 1.37 and $1.42\,M_\odot$ and final total stellar masses of 1.46 and $1.53\,M_\odot$, respectively  --- see Figure~\ref{fig:profile}.
For a symmetric SN producing PSR~J0453+1559, the pre-SN orbital period must be $P_{\rm orb,0}=3.23\,{\rm days}$ (given that the present\footnote{The decay of the orbit due to gravitational wave damping since the formation of the PSR~J0453+1559 system is negligible given its relatively large orbital period.} observed orbital period of the system is $P_{\rm orb}=4.07\,{\rm days}$), whereas the models quoted above were computed with initial orbital periods of 2.0~days, yielding somewhat smaller $P_{\rm orb,0}=1.6-1.8\,{\rm days}$. However, rerunning one of the Tauris~et~al. models with adjusting the companion mass to $1.559\,M_\odot$ and the initial orbital period to 3.0~days (keeping all input physics of the code unchanged), yields a pre-SN orbital period of 3.17~days and an exploding star of total mass $1.59\,M_\odot$.
Thus, further below, we consider the kinematic effects by mimicking tECSNe of stars with total masses between $1.46-1.59\,M_\odot$ (and ONeMg cores of $\la 1.4\,M_\odot$).

\begin{figure}[]
\begin{center}
\mbox{\includegraphics[width=0.7\columnwidth, angle=-90]{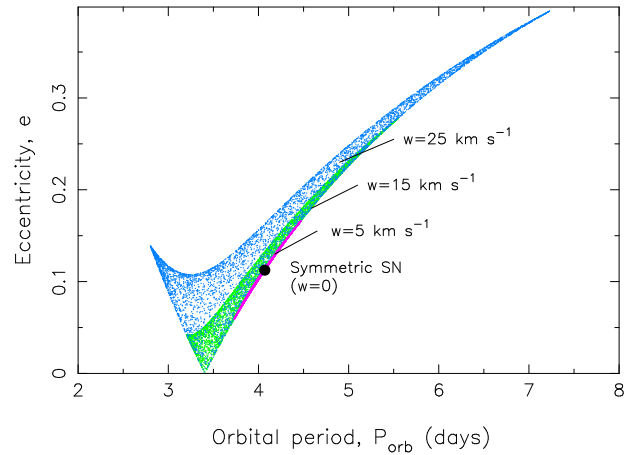}}
  \caption{
    Simulation of post-SN systems in the orbital period--eccentricity plane, based on the properties of PSR~J0453+1559 and an ultra-stripped progenitor star for the tECSN scenario (see text). The result of a symmetric SN corresponds here to the presently observed properties of PSR~J0453+1559. Four kick velocity magnitudes (colour coded: $w = 0$, 5.0, 15.0 and $25.0\,{\rm km\,s}^{-1}$) were each applied in 4000 SNe with random (isotropic) kick directions to illustrate the kinematic effect of even a small kick magnitude. 
    }
\label{fig:kick}
\end{center}
\end{figure}

The effect of a symmetric tECSN producing PSR J0453+1559 results in a 3D systemic recoil velocity of about $12\,{\rm km\,s}^{-1}$. However, proper motion observations of the system \citep{msf+15} point to a likely 3D systemic velocity of order $36-85\,{\rm km\,s}^{-1}$ \citep[corrected for Galactic rotation and location of the binary,][]{tkf+17}, depending on the exact distance to the source and unknown projection of the velocity vector into the plane of the sky. The remaining contribution to the systemic velocity could come either from the first SN of the system, or as a result of a kick during the tECSN. 

With respect to the former possibility, simulations \citep[M.~Kruckow, priv.~comm., see also][]{ktl+18} show that an additional contribution to the systemic velocity of $\sim\!25\,{\rm km\,s}^{-1}$ is not uncommon. In particular, given the large mass of the recycled NS ($1.559\,M_\odot$) in this system, it is likely that the first SN produced a fairly large NS kick at birth \citep{tkf+17,jan17}, resulting in a significant systemic velocity, in agreement with observations of the proper motion. 

\begin{figure*}[t]
\begin{center}
\mbox{\includegraphics[width=0.7\textwidth, angle=-90]{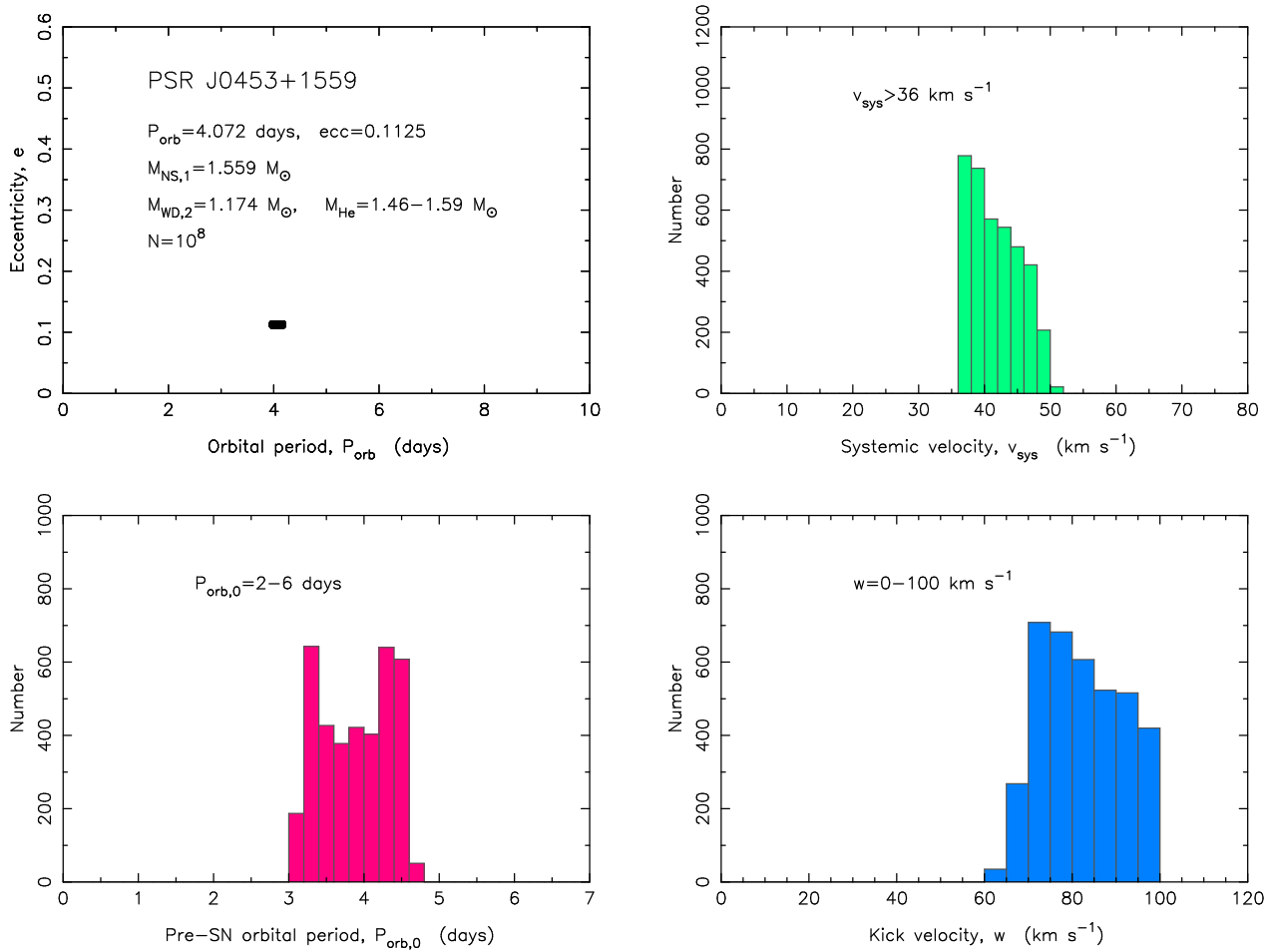}}
  \caption{
      Formation of the PSR~J0453+1559 system if composed of a NS and an ONeFe~WD, based on Monte Carlo simulations of the kinematics of 100~million tECSNe, following the method of \citet{tkf+17}. The four panels display distributions of (upper left to lower right): post-SN orbital period and eccentricity, post-SN 3D systemic velocity, pre-SN orbital period, and magnitude of the kick.
      The total masses of the tECSN progenitor stars at the onset of explosion were chosen from a flat distribution between $1.46-1.59\,M_\odot$, and applied kick velocity magnitudes were chosen from a flat distribution between $0-100\,{\rm km\,s}^{-1}$. The resulting systemic velocities from these computations are in accordance with observational data of PSR~J0453+1559 if the systemic velocity originates exclusively from a kick imparted on the ONeFe~WD in an asymmetric tECSN explosion (see text).
      }
\label{fig:sim_0453_tECSN}
\end{center}
\end{figure*}

For the latter possibility, a kick of at least $69\,{\rm km\,s}^{-1}$ is required to obtain a minimum systemic recoil velocity of $36\,{\rm km\,s}^{-1}$ (see below for details), but only if the estimated systemic velocity would originate exclusively from a kick. More work on tECSN is needed to quantify such possible kicks. A dipole asymmetry $a_\mathrm{d}$ of the radial ejecta momentum, i.e. 
$p_{\mathrm{ej},r}(\theta) = p_{\mathrm{ej},0}(1+a_\mathrm{d}\cos\theta)$, leads to a kick of
the compact remnant of the explosion (a WD for the tECSN case, but similarly for a NS in the 
case of a core-collapse SN) of $w = \alpha_\mathrm{ej} M_\mathrm{ej} M_\mathrm{rem}^{-1}
v_\mathrm{ej}$, where $M_\mathrm{ej}v_\mathrm{ej}$ is the radial ejecta momentum for ejecta mass
$M_\mathrm{ej}$ with average velocity $v_\mathrm{ej}$, and
\begin{equation}
\alpha_\mathrm{ej} = \frac{\int_{-1}^{+1}\mathrm{d}\cos\theta\;(1+a_\mathrm{d}\cos\theta)\cos\theta}{\int_{-1}^{+1}\mathrm{d}\cos\theta\;(1+a_\mathrm{d}\cos\theta)} = \frac{1}{3}\,a_\mathrm{d}
\label{eq:alpha}
\end{equation}
is the relevant ejecta asymmetry parameter. In order to obtain a kick of $\sim\!70\,{\rm km\,s}^{-1}$,
for a WD of 1.17\,$M_\odot$ and an ejecta mass of $\sim$0.3\,$M_\odot$ 
expelled with an average speed of $10^4\,{\rm km\,s}^{-1}$, one needs an ejecta asymmetry $\alpha_\mathrm{ej}$ of about 2.7\,\%, 
corresponding to a dipole moment of the ejecta momentum of $a_\mathrm{d} = 0.08$.
In view of the considerable ejecta asymmetry produced in the oxygen deflagration simulations
of \citet{jrp+16} and \citet{jon+19}, such a magnitude of the dipole component of the radial
momentum distribution appears to be well within reach.

Inferring the precise progenitor system values of PSR~J0453+1559 given the current post-SN data is difficult, even in cases where only a small kick was at work during the formation of the second-born compact object (being a NS or an ONeFe~WD). We illustrate in Figure~\ref{fig:kick} the effects of a small kick applied to the second-formed compact object in PSR~J0453+1559. This model takes its basis in a $1.48\,M_\odot$ exploding star, leaving behind a remnant with a mass of $1.174\,M_\odot$. The pre-SN orbital period is 3.23~days and the mass of the companion star (the recycled NS) is $1.559\,M_\odot$. With these input values, a symmetric SN will leave behind a system with orbital parameters similar to PSR~J0453+1559, whereas a small kick of $25\,{\rm km\,s}^{-1}$ is seen to produce systems with $0<e<0.4$, depending on the kick direction. Applying larger kicks would lead to a wider range of post-SN eccentricities and possibly disruption of the system.

As pointed out above, indeed a larger kick is needed {\em if} the systemic velocity of PSR~J0453+1559 ($\ga 36\,{\rm km\,s}^{-1}$) shall be explained exclusively by the second SN (both in the case of a tECSN as well as a core-collapse SN). This is demonstrated in Figure~\ref{fig:sim_0453_tECSN}, where we have simulated the kinematic effects of 100~million explosions mimicking tECSNe. We sampled the outcome by choosing tECSN explosions of stars of $1.46-1.59\,M_\odot$ and applying kicks between $0-100\,{\rm km\,s}^{-1}$, whereas the companion star mass was fixed at $1.559\,M_\odot$, equal to the mass of the recycled pulsar in the J0453+1559 system. In this case, we find solutions only for $w>69\,{\rm km\,s}^{-1}$. Here a solution refers to a system which has similar orbital period and eccentricity as PSR~J0453+1559 within an error margin of 3\,\% and a systemic velocity compatible with the observational constraints for that binary system.

\section{Discussion and conclusions}\label{sec:conclusions}
The possibility of tECSNe instead of gravitational-collapse ECSNe with NS formation
is controversial \citep{suz+19,zha+19}. It depends on the central density of
the degenerate ONeMg core at oxygen ignition, which again depends on the core growth
rate, convection and semiconvection, and the relevant microphysics such as electron
capture rates and Coulomb effects \citep{zha+19,suz+19,kir+19}. Moreover, it is
unclear whether the bound ONeFe remnant can have a mass of 1.17\,$M_\odot$ after  
a few $0.1\,M_\odot$ have been explosively ejected. Although cases with 
bound remnants of $1.2-1.3\,M_\odot$ and ejecta masses of $\sim\!0.2\,M_\odot$
and $\sim\!0.1\,M_\odot$, respectively, are presented
by \citet{jrp+16,jon+19}, their better resolved 3D simulations suggest bound-remnant 
masses around $0.25-0.4\,M_\odot$ and ejecta masses around $1-1.15\,M_\odot$
\citep[see also][]{kir+19}. However, the modeling of the latest evolution stages and
of the final fate of $\sim\! 8-10\,M_\odot$ stars with strongly degenerate ONeMg 
cores remains highly uncertain, because it depends sensitively on disputed input physics: minimum electron fraction, relative chemical mix of O/Ne/Mg, the uncertain mass-accretion rate of the degenerate ONeMg core, and on the ignition density, initial O-flame structure, and treatment of the oxygen deflagration.
In view of these substantial uncertainties, our hypothetical formation scenario of a 
NS--WD binary with a $1.17\,M_\odot$ WD and orbital eccentricity of $e\sim 0.11$ 
still appears as an interesting, though speculative, possibility for J0453+1559. 

We emphasize that we have previously demonstrated \citep{tkf+17} that all kinematic properties, as well as the spin and B-field, of the observed recycled pulsar can be well accounted for by a low-mass core-collapse SN producing a double NS system, {\em assuming} the possibility that a $1.17\,M_\odot$ NS can be produced in a stellar core collapse. However, here we have argued that this assumption is by no means assured, and therefore we present an alternative formation hypothesis based on the tECSN scenario

More work on tECSNe and their stellar progenitors is needed to assess the viability 
of this scenario. As long as the open problems of the tECSN phenomenon as a
channel of WD formation are not settled, arguments that classify PSR~J0453+1559
as a double NS system on grounds of its large orbital eccentricity alone cannot be 
considered as rock-solid.

An additional open question is whether or not other cases of compact binaries classified as double NS systems could possibly have formed via the investigated tECSN scenario, too.
PSR~J0453+1559 has a companion mass of $1.174(4)\,M_\odot$ which is significantly lower than that of the candidate Galactic double NS system with the second lowest mass, PSR~J1756$-$2251 with $1.230(7)\,M_\odot$ \citep{fsk+14}.
The latter could potentially be produced via an iron-core collapse SN or a ``normal'' ECSN of a collapsing progenitor. The possible coexistence and differences between normal ECSNe and tECSNe also requires further investigation. 
Upcoming measurements of NS masses from anticipated new discoveries of binary radio pulsars by the Square Kilometre Array \citep[SKA,][]{kbs+15}, and detection of additional double NS mergers with LIGO \citep{aaa+17}, will constrain the minimum NS mass further and shed light on the formation paths of binary NSs as well as the final stages of stellar evolution and SN physics.

\acknowledgements
We are grateful to R\"udiger Pakmor for information on tECSN explosions and uncertainties in their modeling.
T.M.T.\ acknowledges an AIAS--COFUND Senior Fellowship funded by the European Union’s Horizon~2020 Research and Innovation Programme (grant agreement no~754513) and Aarhus University Research Foundation.
At Garching, funding by the
European Research Council through grant ERC-AdG No.~341157-COCO2CASA
and by the Deutsche Forschungsgemeinschaft through grants
SFB-1258 ``Neutrinos and Dark Matter in Astro- and Particle Physics
(NDM)'' and EXC~2094 ``ORIGINS: From the Origin
of the Universe to the First Building Blocks of Life'' is acknowledged.

\bibliography{references}
\bibliographystyle{aasjournal}

\end{document}